\documentclass[preprint,aps,12pt,showpacs,nofootinbib,tightenlines]{revtex4}

\usepackage[section]{placeins}
\usepackage{graphicx}
\usepackage{dcolumn}
\usepackage{bm}

\usepackage{epsfig}
\usepackage{amssymb}
\usepackage{amsmath}
\usepackage{flafter}
\usepackage{array}

\usepackage[dvips,usenames]{color}

\newlength{\dinwidth}
\newlength{\dinmargin}
\newcommand{\spur}[1]{\not\! #1 \,}
\begin{document}
\title{QCD Approach to $B\to D\pi$ Decays and CP Violation}
\author{Fang Su$^{a,b}$,  Yue-Liang Wu$^a$, Ya-Dong Yang$^c$,
and Ci Zhuang$^{a,b}$} \affiliation{$^a$Kavli
Institute for Theoretical Physics China, Institute of Theoretical
Physics \\ Chinese Academy of Science (KITPC/ITP-CAS), Beijing
100080, China \\ $^b$Graduate School of the Chinese Academy of
Science, Beijing, 100039, China\\ $^c$Department of Physics, Henan
Normal University, Xinxiang, Henan 453007, China }

\begin{abstract}
\noindent The branching ratios and CP violations of the $B\to D\pi$
decays, including both the color-allowed and the color-suppressed
modes, are investigated in detail within QCD framework by
considering all diagrams which lead to three effective currents of
two quarks. An intrinsic mass scale as a dynamical gluon mass is
introduced to treat the infrared divergence caused by the soft
collinear approximation in the endpoint regions, and the Cutkosky
rule is adopted to deal with a physical-region singularity of the on
mass-shell quark propagators. When the dynamical gluon mass $\mu_g$
is regarded as a universal scale, it is extracted to be around
$\mu_g = 440$ MeV from one of the well-measured $B\to D\pi$ decay
modes. The resulting predictions for all branching ratios are in
agreement with the current experimental measurements. As these
decays have no penguin contributions, there are no direct $CP$
asymmetries. Due to interference between the Cabibbo-suppressed and
the Cabibbo-favored amplitudes, mixing-induced $CP$ violations are
predicted in the $B\to D^{\pm}\pi^{\mp}$ decays to be consistent
with the experimental data at 1-$\sigma$ level. More precise
measurements will be helpful to extract weak angle $2\beta+\gamma$.
\end{abstract}

\maketitle

\section{Introduction}

Nonleptonic $B$-meson decays are of crucial importance to deepen
our insights into the flavor structure of the Standard Model~(SM),
the origin of CP violation, and the dynamics of hadronic decays,
as well as to search for any signals of new physics beyond the SM.
However, due to the non-perturbative strong interactions involved
in these decays, the task is hampered by the computation of matrix
elements between the initial and the final hadron states. In order
to deal with these complicated matrix elements reliably, several novel
methods based on the naive factorization
approach~(FA)~\cite{Wirbel}, such as the QCD factorization
approach~(QCDF)~\cite{M}, the perturbation QCD
method~(pQCD)~\cite{lihn}, and the soft-collinear effective
theory~(SCET)~\cite{SCET}, have been developed in the past few
years. These methods have been used widely to analyze the hadronic
$B$-meson decays, while they have very different understandings
for the mechanism of those decays, especially for the case of
heavy-light final states, such as the $B\to D\pi$ decays.
Presently, all these methods can give good predictions for the
color allowed $\overline B^0\to D^+\pi^-$ mode, but for the color
suppressed $\overline B^0\to D^0\pi^0$ mode, the QCDF and the SCET
methods could not work well, and the pQCD approach seems leading
to a reasonable result in comparison with the experimental data.
In this situation, it is interesting to study various approaches
and find out a reliable approach.

As the mesons are regarded as quark and anti-quark bound states, the
nonleptonic two body meson decays concern three quark-antiquark
pairs. It is then natural to investigate the nonleptonic two body
meson decays within the QCD framework by considering all Feynman
diagrams which lead to three effective currents of two quarks. In
our considerations, beyond these sophisticated pQCD, QCDF and SCET,
we shall try to find out another simple reliable QCD approach to
understand the nonleptonic two body decays. In this note, we are
focusing on evaluating the $B\to D\pi$ decays.

The paper is organized as follows. In Sect. II, we first analyze the
relevant Feynman diagrams and then outline the necessary ingredients
for evaluating the branching ratios and $CP$ asymmetries of $B\to
D\pi$ decays. In Sect. III, we list amplitudes of $B\to D\pi$
decays. The approaches for dealing with the physical-region
singularities of gluon and quark propagators are given in Sect. IV.
Finally, we discuss the branching ratios and the $CP$ asymmetries
for those decay modes and give conclusions in Sects. V and VI,
respectively. The detail calculations of amplitudes for these decay
modes are given in the Appendix.

\section{Basic Considerations for Evaluating $B \to D\pi$ Decays}

We start from the four-quark effective operators in the effective
weak Hamiltonian, and then calculate all the Feynman diagrams which
lead to effective six-quark interactions. The effective Hamiltonian
for $\overline B\to D\pi$ decays can be expressed as
\begin{equation}\label{Heff}
{\cal
H}_{eff}=\frac{G_{F}}{\sqrt{2}}V_{cb}V_{ud}^*[C_{1}(\mu)O_{1}(\mu)+
C_{2}(\mu)O_{2}(\mu)]+{\rm h.c.},
\end{equation}
where $C_1$ and $C_2$ are the Wilson coefficients which have been
evaluated at next-to-leading order~\cite{Buchalla}, $O_1$ and $O_2$
are the tree operators arising from the $W$-boson exchanges with
\begin{equation}
\begin{array}{ll}
O_{1}=(\bar{c}_{i}b_{i})_{V-A}(\bar{d}_{j}u_{j})_{V-A},
\hspace{1.5cm} &
O_{2}=(\bar{c}_{i}b_{j})_{V-A}(\bar{d}_{j}u_{i})_{V-A},
\end{array}
\end{equation}
where $i$ and $j$ are the ${\rm SU(3)}$ color indices.

\begin{figure}[t]
\begin{center}
\scalebox{0.9}{\epsfig{file=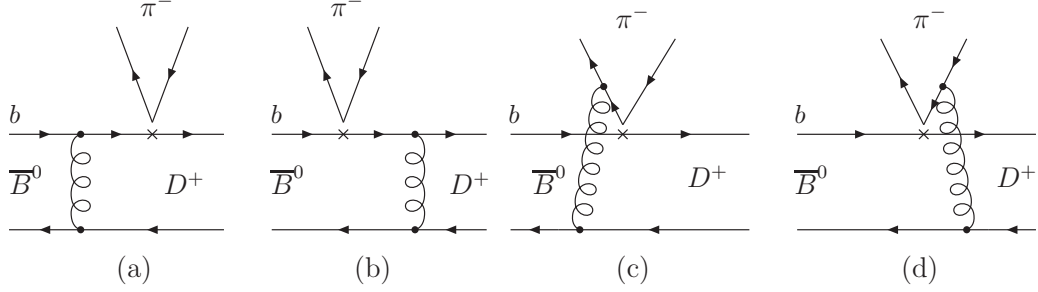}} \caption{\label{tree} \small
The factorizable~((a) and (b)) and nonfactorizable~((c) and (d))
diagrams contributing to the color-allowed $\overline
{B}^0\rightarrow D^{+}\pi^{-}$ decay.}
\end{center}
\end{figure}

\begin{figure}[t]
\begin{center}
\scalebox{0.9}{\epsfig{file=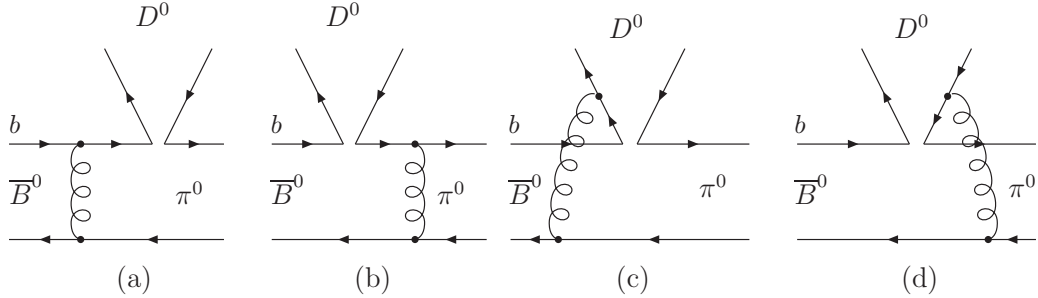}} \caption{\label{tree2}
\small The factorizable~((a) and (b)) and nonfactorizable~((c) and
(d)) diagrams contributing to the color-suppressed $\overline
{B}^0\rightarrow D^{0}\pi^{0}$ decay.}
\end{center}
\end{figure}

\begin{figure}[t]
\begin{center}
\scalebox{0.7}{\epsfig{file=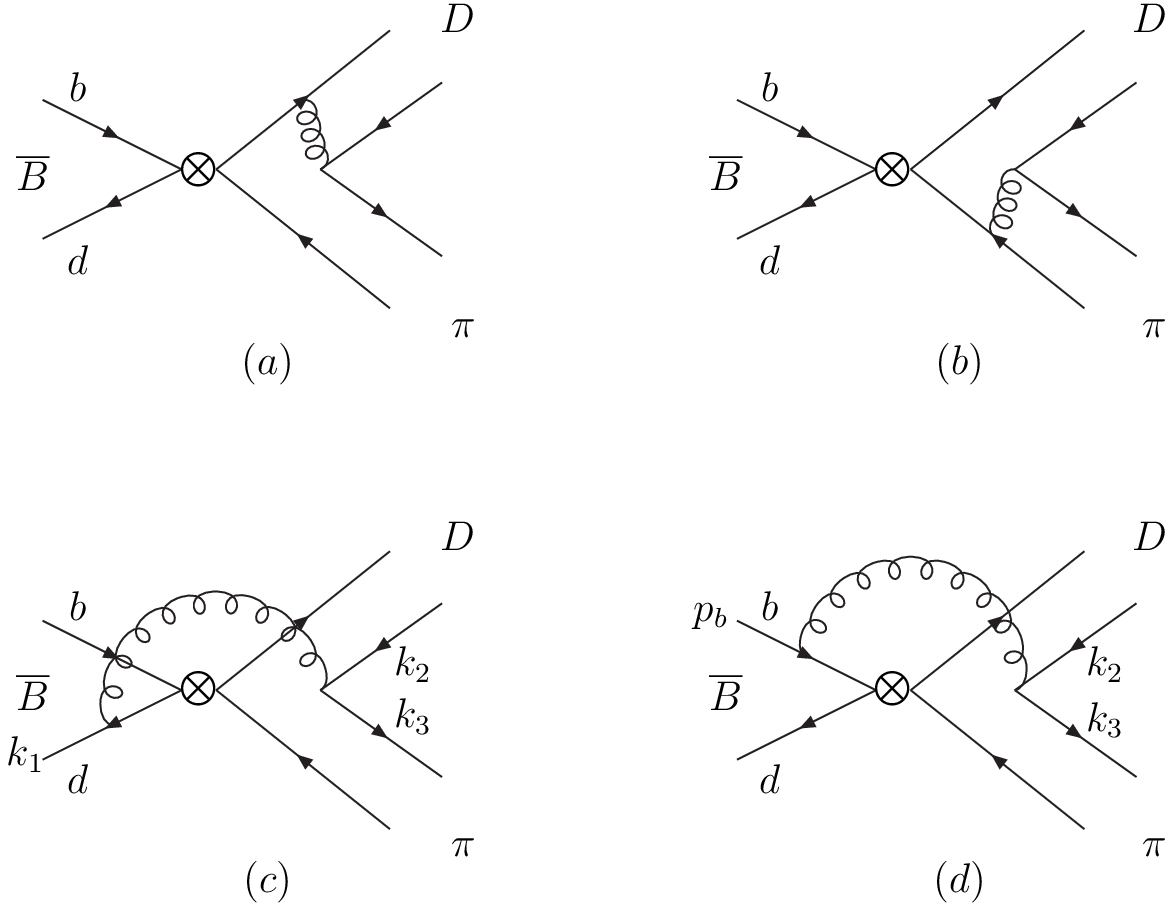}}
\caption{\label{annihilation} \small The annihilation diagrams for
$\overline{B}^0\rightarrow D^{+}\pi^{-}$ and $\overline
{B}^0\rightarrow D^{0}\pi^{0}$ decays.}
\end{center}
\end{figure}

Based on the effective Hamiltonian in Eq.~(\ref{Heff}), we can then
calculate the decay amplitudes for $\overline B^0\rightarrow
D^+\pi^-$, $\overline B^0\rightarrow D^0\pi^0$, and $B^-\rightarrow
D^0\pi^-$ decays, which are the color-allowed, the color-suppressed,
and the color-allowed plus color-suppressed modes, respectively. All
the six-quark Feynman diagrams that contribute to $\overline
B^0\rightarrow D^+\pi^-$ and $B^0\rightarrow D^0\pi^0$ decays are
shown in Figs.~\ref{tree}-\ref{annihilation} via one gluon exchange.
As for the process $B^-\rightarrow D^0\pi^-$, it doesn't involve the
annihilation diagrams and the related Feynman diagrams are the sum
of Figs.~\ref{tree} and \ref{tree2}. Based on the Isospin symmetry
argument, the decay amplitude of this mode can be written as
$A(B^-\rightarrow D^0\pi^-)=A(\overline B^0\rightarrow
D^+\pi^-)-\sqrt{2}A(\overline B^0\rightarrow D^0\pi^0)$. The
explicit expressions for the amplitudes of these decay modes are
given in detail in next section.


The decay amplitudes of $B \to D\pi$ decay modes are quite
different. For the color-allowed $\overline B^0\to D^+\pi^-$ mode,
it is expected that the decay amplitude is dominated by the
factorizable contribution $A_{fac}$~(from the diagrams (a) and (b)
in Fig.~\ref{tree}), while the nonfactorizable contribution
$A_{nonfac}$~(from the diagrams (c) and (d) in Fig.~\ref{tree})
has only a marginal impact. This is due to the fact that the
former is proportional to the large coefficient
$a_1=C_1+\frac{C_2}{N_C}\sim 1$, while the latter is proportional
to the quite small coefficient $a_2=C_2+\frac{C_1}{N_C}\sim 0$. In
addition, there is an addition color-suppressed factor
$\frac{1}{N_C}$ in the nonfactorizable contribution $A_{nonfac}$.
In contrast with the $\overline B^0\to D^+\pi^-$ mode, the
nonfactorizable contribution $A_{nonfac}$~(from (c) and (d)
diagrams in Fig.~\ref{tree2}) in the $\overline B^0\to D^0\pi^0$
mode is proportional to the large coefficient
$a_1=C_1+\frac{C_2}{N_C}\sim 1$, and even if with an additional
color-suppressed factor $\frac{1}{N_C}$, its contribution is still
larger than the factorizable one $A_{fac}$~(from (a) and (b)
diagrams in Fig.~\ref{tree2}) which is proportional to the quite
small coefficient $a_2=C_2+\frac{C_1}{N_C}\sim 0$. Thus, it is
predicted that the decay amplitude of this mode is dominated by
the nonfactorizable contribution $A_{nonfac}$. As for the $B^- \to
D^0\pi^-$ mode, since its amplitude can be written as the sum of
the ones of the above two modes, it is not easy to see which one
should dominate the total amplitude.

The branching ratio for $B\to D\pi$ decays can be expressed as
follows in terms of the total decay amplitudes
\begin{eqnarray}
{\cal B}(B \to D\pi)=\frac{\tau_B \,p_c}{8\,\pi\,
m_B^2}\,\left|{\cal A}(B\to D\pi)\right|^2 \,
\end{eqnarray}
where $\tau_B$ is the lifetime of the $B$ meson, and $p_c$ is the
magnitude of the momentum of the final-state particles $D$ and $\pi$
in the $B$-meson rest frame and given by
\begin{eqnarray}
p_c=\frac{1}{2m_B}\sqrt{\left[m_B^2-(m_{D}+m_{\pi})^2\,\right]\,
\left[m_B^2-(m_{D}-m_{\pi})^2\,\right]}\,.
\end{eqnarray}

As is well-known, the direct $CP$ violation in meson decays is
non-zero only if there are two contributing amplitudes with non-zero
relative weak and strong phases. The weak-phase difference usually
arises from the interference between two different topological
diagrams. For three $B\to D\pi$ decays, it is seen from the Feynman
diagrams in Figs.~\ref{tree}-\ref{annihilation} that there are no
weak-phase differences, and hence no direct $CP$ violation in all
these three modes, we shall then consider the mixing-induced $CP$
violation.

As the final states $D^+\pi^-$ can be produced both in the decays of
$\overline B^0$ meson via the Cabibbo-favored~($b\to c$) and in the
decays of $B^0 $ meson via the doubly Cabibbo-suppressed~($b\to u$)
tree amplitudes. The relative weak-phase difference between these
two amplitudes is $-\gamma$ and, when combining with the
$B^0-\overline B^0$ mixing phase, the total weak-phase difference is
$-(2\beta+\gamma)$ to all orders in the small CKM parameter
$\lambda$. Thus, the $B^0(\overline B^0)\to D^+\pi^-$ decays can in
principle be used to measure the weak phase $\gamma$, since the weak
phase $\beta$ has been measured with high precision. The
time-dependent $CP$ asymmetry of such decay modes is defined as:
\begin{eqnarray}
{\cal  A}_{D^+\pi^-}(\Delta t) &=& \frac{\Gamma (\overline B^0 \to
D^+\pi^-(\Delta t)) -\Gamma( {B}^0 \to D^+\pi^-(\Delta t))}{ \Gamma
(\overline B^0\to D^+\pi^-(\Delta t))+\Gamma( {B}^0 \to
D^+\pi^-(\Delta t))}\nonumber\\
& \simeq & -a_{\epsilon} + (a_{\epsilon} +
a_{\epsilon'}^{D^+\pi^-})\cos{(\Delta m_B\cdot t)} + a_{\epsilon +
\epsilon'}^{D^+\pi^-}
\sin{(\Delta m_B\cdot t)} \nonumber \\
 &\simeq & S_{D^+\pi^-}\sin(\Delta mt)-C_{D^+\pi^-}\cos(\Delta
mt),\label{eq:acp0}
\end{eqnarray}
where $\Delta m$ is the mass difference of the two eigenstates of
$B_d$ mesons, and $S_{D^+\pi^-}$ and $C_{D^+\pi^-}$ are given as
\begin{equation}
S_{D^+\pi^-}=\frac{2Im(\lambda_{D^+\pi^-})}{1+|\lambda_{D^+\pi^-}|^2}
= a_{\epsilon + \epsilon'}^{D^+\pi^-} , \hspace{1cm}
C_{D^+\pi^-}=\frac{1-|\lambda_{D^+\pi^-}|^2}{1+|\lambda_{D^+\pi^-}|^2}=
-(a_{\epsilon} + a_{\epsilon'}^{D^+\pi^-}),
\end{equation}
with
\begin{equation}
\lambda_{D^+\pi^-}=\frac{V^*_{tb}V_{td}\langle
D^+\pi^-|H_{eff}|\overline B^0\rangle}{V_{tb}V^*_{td}\langle
D^+\pi^-|H_{eff}|B^0\rangle}.
\end{equation}
Where the rephase-invariant quantities $a_{\epsilon}$,
$a_{\epsilon'}$ and $a_{\epsilon + \epsilon'}$ \cite{PW}
characterize the indirect, direct and mixing-induced CP violations
respectively. As $a_{\epsilon}\ll 1$ for neutral $B$ system, we have
$C_{D^+\pi^-} \simeq - a_{\epsilon'}^{D^+\pi^-}$ which characterizes
direct CP violation. Defining $g$ and $h$ as the amplitudes of
$\overline B^0\rightarrow D^+\pi^-$ and $B^0\rightarrow D^+\pi^-$
decay modes, respectively, we can further express these two CP
asymmetries as
\begin{eqnarray}
C_{D^+\pi^-}&=&\frac{|h|^2-|g|^2}{|h|^2+|g|^2}=\frac{z^2-1}{z^2+1} \\
S_{D^+\pi^-}&=&\frac{-2|g||h|\sin(2\beta+\gamma+\delta)}{|h|^2+|g|^2}
=\frac{-2z\sin(2\beta+\gamma+\delta)}{z^2+1},\label{cs1}
\end{eqnarray}
where $z=|h/g|$, and $\delta={\rm arg}|h/g|$ represents the relative
strong-phase difference between the two amplitudes $g$ and $h$.

Similarly, we can define another two $CP$-violating parameters
$C_{D^-\pi^+}$ and $S_{D^-\pi^+}$ for the $B^0(\overline B^0)\to
D^-\pi^+$ decays
\begin{eqnarray}
C_{D^-\pi^+}=\frac{1-\bar{z}^2}{1+\bar{z}^2},\hspace{1cm}
S_{D^-\pi^+}=\frac{-2\bar{z}\sin(2\beta+\gamma-\delta)}{1+\bar{z}^2},\label{cs2}
\end{eqnarray}
with the parameter $\bar{z}$ defined as $\bar{z} =
|\bar{h}/\bar{g}|=z$. Here the amplitudes $\bar{h}$ and $\bar{g}$
are the charge conjugations of the amplitudes $h$ and $g$. Since the
magnitude of the Cabibbo-suppressed decay amplitude $|h|$ is much
smaller than that of the Cabibbo-favored decay amplitude $|g|$, the
ratio $z$ should be quite small and is found to be about $0.02$ in
our framework. Thus, to a very good approximation,
$C_{D^+\pi^-}=-C_{D^-\pi^+}\simeq -1$, and the coefficients of the
sine terms are given by
\begin{equation}
S_{D^+\pi^-}=-2z\sin(2\beta+\gamma+\delta)=  a_{\epsilon +
\epsilon'}^{D^+\pi^-}\,, \hspace{1cm}
S_{D^-\pi^+}=-2z\sin(2\beta+\gamma-\delta) =  a_{\epsilon +
\epsilon'}^{D^-\pi^+}.
\end{equation}
To compare with the current experimental data, one usually define
the following two quantities, which are given by the combination of
two $CP$-violating parameters $S_{D^+\pi^-}$ and $S_{D^-\pi^+}$,
\begin{equation}
a=(S_{D^+\pi^-}+S_{D^-\pi^+})/2\,,\hspace{1cm}c=(S_{D^+\pi^-}-S_{D^-\pi^+})/2,
\label{cs3}
\end{equation}
which can provide constraints on the weak phase $2\beta+\gamma$ and
the strong phase $\delta$.

\section{$B\to D\pi$ Decay Amplitudes}

Using the methods given in the Appendix, we can get the $B\to D\pi$
decay amplitudes, which are composed of three parts: the
factorizable contribution $A_{fac}$, the nonfactorizable
contribution $A_{nonfac}$, and the annihilation contribution
$A_{anni}$. The amplitude of $\overline B^0\to D^+\pi^-$ mode is
found to be
\begin{equation}
{\cal A}(\overline B^0\to
D^{+}\pi^{-})=V_{cb}V^*_{ud}(A_{fac}+A_{nonfac}+A_{anni}),\label{favor1}
\end{equation}
with
\begin{eqnarray}
A_{fac}&=&\frac{G_{F}} {\sqrt{2}}f_{B}f_Df_\pi
\pi\alpha_s(\mu)(C_1+\frac{C_2}{N_C})\frac{C_F}{N_C}
\int_0^1dx\int_0^1dy\, \phi_B(x)\phi_D(y) \nonumber\\& \times &
\biggl((-\bar y m_B^2+2 m_b m_B-2\bar y m_D m_B + m_b
m_D)\frac{m_B^2}{D_bk^2}\nonumber\\
&+&(-m_c m_B-2\bar x m_D m_B -2m_c m_D)\frac{m_B^2}{D_c
k^2}\bigg),\,\label{favor2}
\end{eqnarray}
where $D_b=m^2_B\bar y-m^2_b $, $D_c=-m^2_B x\bar x-m^2_c $ and
$k^2=m^2_B x(x-y)$. $\phi's$ are the wave functions of mesons. For
the $B$-meson wave function, we shall take the form given
in~\cite{h.y.cheng}
\begin{equation}
\phi_B( \rho)=N_B \rho^2(1-\rho)^2
\textrm{exp}\left[-\frac{1}{2} \left(\frac{ \rho
m_B}{\omega_B}\right)^2\right],
\end{equation}
with $\omega_B=0.25~{\rm GeV}$, and $N_B$ being a normalization
constant. The $D$ meson distribution amplitude is given by
\begin{equation}
\phi_D(y)=6y(1-y)[1+C_D(1-2y)],
\end{equation}
with the shape parameter $C_D=0.8$. For the $\pi$ meson light cone
wave functions, we use the asymptotic form as given in
Refs.~\cite{M.T,Genon,p.ball}:
\begin{eqnarray}
\phi(u)=\phi_\sigma(u)=6u\bar u,\hspace{1cm}\phi_\pi(u)=1.
\end{eqnarray}
with $\bar u=1-u$.

\begin{eqnarray}
A_{nonfac}&=&\frac{G_{F}} {\sqrt{2}}f_{B}f_Df_\pi
\pi\alpha_s(\mu)(C_2+\frac{C_1}{N_C})\frac{C_F}{N_C^2}
\int_0^1dx\int_0^1dy\int_0^1dz\, \phi_B(x)\phi_D(y)\phi(z)\nonumber\\
&\times & \bigg(\big((-x+z)m_B^2 -(x-y) m_D m_B\big)\frac{m_B^2}{D_d k^2}\nonumber\\
&+&\big((-2x + y +\bar z)m_B^2-(x-y)m_D m_B\big)\frac{m_B^2}{D_u
k^2}\biggl)\,,\label{favor3}
\end{eqnarray}
where $D_d=m^2_B(x-y)(x-\bar z)$ and $D_u=m^2_B(x-y)(x-z)$.
\begin{eqnarray}
A_{anni}&=&\frac{G_{F}} {\sqrt{2}}f_{B}f_D f_\pi
\pi\alpha_s(\mu)\frac{C_F}{N_C^2}
\int_0^1dx\int_0^1dy\int_0^1dz\,\phi_D(y)\biggl\{\nonumber\\
&\times&\biggl[(C_2+\frac{C_1}{N_C})\biggl((\bar z
m^2_B-2m_cm_D)\phi(z)+\mu_\pi(
m_c-2zm_D+4m_D)\phi_\pi(z)\nonumber\\&-&
\mu_\pi(m_c+2zm_D)\frac{\phi_\sigma'(z)}{6}
\biggl)\frac{m_B^2}{D_{ca} k_a^2}+\biggl(-ym^2_B\phi(z)+2\mu_\pi
m_D(y+1)\phi_\pi(z)\biggl)\frac{m_B^2}{D_{ua} k_a^2} \biggl]\nonumber\\
&+&(C_1+\frac{C_2}{N_C})\phi_B(x)\biggl[\biggl(\big((\bar
x-y)m_B+m_b \big)m^2_B\phi(z)+ \mu_\pi m_D \big((2x-\bar y-z)m_B
\nonumber\\&-&4m_b \big)\phi_\pi(z) +\mu_\pi m_Dm_B(y-\bar
z)\frac{\phi_\sigma'(z)}{6}\biggl)
\frac{m_B}{D_{ba} k_a^2}\nonumber\\
&+&\biggl((x-\bar z)m^2_B\phi(z)-\mu_\pi m_D\big((y-\bar
z)\frac{\phi_\sigma'(z)}{6}+ (2x-y-\bar
z)\phi_\pi(z)\big)\biggl)\frac{m^2_B}{D_{da}
k_a^2}\biggl]\biggl\},\label{anni1}
\end{eqnarray}
where $\phi_\sigma'(z)=\frac{d\phi_\sigma(z)}{dz}$, $D_{ca}=m^2_B
z-m^2_c$, $D_{ua}=m^2_B \bar y$, $D_{ba}=m^2_B(\bar x-y)(\bar
x-z)-m^2_b$, $D_{da}=m^2_B(x-y)(x-z)$ and $k^2_a=m^2_B yz$. The
annihilation contribution is found to be much smaller than the ones
from the factorizable and the nonfactorizable diagrams. Numerically,
it is negligible.

For the color-suppressed $\overline B^0 \to D^0\pi^0$ decay, its
amplitude can be written as
\begin{equation}
{\cal A}(\overline B^0\to
D^0\pi^0)=-\frac{1}{\sqrt{2}}V_{cb}V^*_{ud}(A_{fac}+A_{nonfac}-A_{anni}),
\end{equation}
with
\begin{eqnarray}
A_{fac}&=&\frac{G_{F}} {\sqrt{2}}f_{B}f_Df_\pi
\pi\alpha_s(\mu)(C_2+\frac{C_1}{N_C})\frac{C_F}{N_C}
\int_0^1dx\int_0^1dz\, \phi_B(x) \nonumber\\& \times &
\bigg[\biggl((-\bar z m_B^2+2 m_b m_B)\phi(z)+ \mu_\pi (2\bar z m_B
-m_b)\phi_\pi(z)\nonumber\\
&+& \mu_\pi \big(2(z+1)m_B + m_b \big)
\frac{\phi_\sigma'(z)}{6}\bigg)\frac{m_B^2}{D_bk^2} + 2\mu_\pi \bar
x \phi_\pi(z)\frac{m_B^3}{D_d k^2} \biggl] ,\label{ckms1}
\end{eqnarray}
here $D_b=m^2_B\bar z-m^2_b$,  $D_d=m^2_B(x-\bar y)(x-z)$ and
$k^2=m^2_B x(x-z)$.
\begin{eqnarray}
A_{nonfac}&=&\frac{G_{F}} {\sqrt{2}}f_{B}f_Df_\pi
\pi\alpha_s(\mu)(C_1+\frac{C_2}{N_C})\frac{C_F}{N_C^2}
\int_0^1dx\int_0^1dy\int_0^1dz\, \phi_B(x)\phi_D(y)\nonumber\\
&\times & \bigg[\biggl(\big((\bar x-y)m_B^2 + m_c m_D \big)
\phi(z)+\mu_\pi m_B (x-z)\biggl(\phi_\pi(z)
-\frac{\phi_\sigma'(z)}{6}\biggl)\biggl)\frac{m_B^2}{D_c k^2}\nonumber\\
&+&\bigg((-2x+y+z)m_B\phi(z)-\mu_\pi(x-z)\biggl(\phi_\pi(z)
-\frac{\phi_\sigma'(z)}{6}\biggl) \bigg)\frac{m_B^3}{D_u
k^2}\biggl]\,,\label{ckms2}
\end{eqnarray}
where $D_c=-m^2_B x\bar x-m^2_c$ and $D_u=m^2_B(x-y)(x-z)$. For the
annihilation amplitude $A_{anni}$, it is the same as the one in
Eq.~(\ref{anni1}) since the two modes $D^0\pi^0$ and $D^+\pi^-$ have
the same annihilation topological diagrams.

For the doubly Cabibbo-suppressed decay mode $\overline B^0\to
D^-\pi^+$, its decay amplitude can be written as
\begin{equation}
{\cal A}(\overline B^0\to
D^-\pi^+)=V_{ub}V^*_{cd}(A_{fac}+A_{nonfac}+A_{anni}),
\end{equation}
here, $A_{fac}, A_{nonfac}$ and $A_{anni}$ can be obtained from the
ones of decay mode $\bar{B}^0\to D^0\pi^0$ by simply exchanging the
Wilson coefficients $C_1$ and $C_2$.

For the $B^-\to D^0\pi^-$ decay, its amplitude can be yielded by
using the isospin relation $A(B^-\rightarrow D^0\pi^-)=A(\overline
B^0\rightarrow D^+\pi^-)-\sqrt{2}A(\overline B^0\rightarrow
D^0\pi^0)$.

\section{Treatments for Physical-region Singularities}

To perform a numerical calculation of the decay amplitudes of $B \to
D\pi$ decays, the light-cone projectors of mesons are found to be
very useful, and the details of these quantities are presented in
the Appendix. Where one encounters the endpoint divergences stemming
from the convolution integrals of the meson distribution amplitudes
with the hard kernels, which is caused by the collinear
approximation. To regulate such an infrared divergence, we may
introduce an intrinsic mass scale realized in the
symmetry-preserving loop regularization\cite{LR,LRC}. At the tree
level, it is equivalent to adopt an effective dynamical gluon mass
in the propagator. Practically, such a gluon mass scale has been
used to regulate the infrared divergences in the soft endpoint
region~\cite{Cornwall,Yang,kk}
\begin{equation}
\frac{1}{k^{2}}~\Rightarrow~\frac{1}{k^{2}-\mu_g^2(k^{2})+i\epsilon}\,,
\hspace{1cm}\mu_g^2(k^{2})=\mu_g^2 \biggl
[\frac{\ln(\frac{k^{2}+4\mu_g^2}{\Lambda^{2}})}
{\ln(\frac{4\mu_g^2}{\Lambda^{2}})}\biggl ]^{-\frac{12}{11}},
\end{equation}
The use of this effective gluon propagator is supported by the
lattice~\cite{Williams} and the field theoretical
studies~\cite{Alkofer}, which have shown that the gluon propagator
is not divergent as fast as $\frac{1}{k^{2}}$. Taking the hadronic
scale $\Lambda = \Lambda_{QCD}$, the dynamical gluon mass scale can
be determined from one of the well measured decay mode. Numerically,
we will see that taking $\Lambda_{\rm QCD}=300$~MeV, the dynamical
gluon mass scale is around $\mu_{g}=(1.5\pm 0.2)\Lambda_{\rm QCD}$.

Another physical-region singularity arises from the on mass-shell
quark propagators. It can be easily checked that each Feynman
diagram contributing to a given matrix element is entirely real
unless some denominators vanish with a physical-region singularity,
so that the $i\epsilon$ prescription for treating the poles becomes
relevant. In other words, a Feynman diagram will yield an imaginary
part for the decay amplitudes only when the virtual particles in the
diagram become on mass-shell, thus the diagram may be considered as
a genuine physical process. The Cutkosky rules~\cite{cutkosky} give
a compact expression for the discontinuity across the cut arising
from a physical-region singularity. When applying the Cutkosky rules
to deal with a physical-region singularity of quark propagators, the
following formula holds
\begin{eqnarray}
\frac{1}{(k_1-k_2-k_3)^2+i\epsilon}&=&P\biggl[\frac{1}{(k_1-k_2-k_3)^2}
\biggl]-i\pi\delta[(k_1-k_2-k_3)^2],\label{quarkd}\\
\frac{1}{(p_b-k_2-k_3)^2-m_b^2+i\epsilon}&=&P\biggl[\frac{1}
{(p_b-k_2-k_3)^2-m_b^2}\biggl]-i\pi\delta[(p_b-k_2-k_3)^2-m_b^2],
\label{quarkb}
\end{eqnarray}
where $P$ denotes the principle-value prescription. The role of the
$\delta$ function is to put the particles corresponding to the
intermediate state on their positive energy mass-shell, so that in
the physical region, the individual Feynman diagram satisfies the
unitarity condition. Equations (\ref{quarkd}) and (\ref{quarkb})
will be applied to the quark propagators $D_{da}$ and $D_{ba}$ in
Equation (\ref{anni1}), respectively. It is then seen that the
possible large imaginary parts arise from the virtual quarks across
their mass shells as physical-region singularities.

\section{Numerical Results}

It is seen that for theoretical predictions it depends on many input
parameters, such as the Wilson coefficient functions, the CKM matrix
elements, the hadronic parameters, and so on. To carry out a
numerical calculation, we take the following input
parameters~\cite{yao}
\begin{equation}
\begin{array}{llll}
C_1 =1.117(1.073), & C_2 =-0.267(-0.179),&m_{B}=5.28~{\rm GeV},&m_D=1.87~{\rm GeV},\\
m_{\pi^\pm}=139.6~{\rm MeV},&m_{\pi^0}=135~{\rm
MeV},&m_{b}=4.66~{\rm GeV}, &m_c=1.47~{\rm
GeV},\\f_{B^0}=216\pm19~{\rm MeV},&f_{D }=223\pm17~{\rm
MeV},&f_{\pi}=130.1~{\rm MeV}, &\tau_{B^0}=1.536~{\rm
ps},\\\tau_{B^-}=1.638~{\rm
ps},&V_{ud}=1-\lambda^2/2,&V_{ub}=A\lambda^3(\rho-i\eta),&V_{cd}=-\lambda,\\
 V_{cb}=A\lambda^2.
\end{array}
\end{equation}
The Wolfenstein parameters of the CKM matrix elements are taken
as~\cite{Charles:2004jd}:
$\lambda=0.2272\pm0.001,~A=0.806\pm0.014,~\bar\rho=0.195^{+0.024}_{-0.067},
~\bar\eta=0.326^{+0.032}_{-0.015}$, with
$\bar\rho=\rho(1-\frac{\lambda^2}{2}),~\bar\eta=\eta(1-
\frac{\lambda^2}{2})$. The coefficient of the twist-3 distribution
amplitude of the pseudoscalar $\pi$ meson is chosen as $\mu_\pi=
1.5\pm0.2~{\rm GeV}$~\cite{M,bn}.

With the above values for the input parameters, we are able to
calculate the contributions of different amplitudes for each decay
mode. Our final results at $m_b/2$ scale are presented in
Table~\ref{amplitude}.

\begin{table}[t]
\caption{\label{amplitude} Numerical results at $m_b/2$ scale of the
amplitudes for different diagrams in $B \to D\pi$ decays. Amplitudes
$A_{fac}, A_{nonfac}$, and $A_{anni}$ represent the
factorizable~((a)and (b) diagrams in Figs.~\ref{tree} or
\ref{tree2}), the non-factorizable~((c) and (d) diagrams in
Figs.~\ref{tree} or \ref{tree2}), and the annihilation~(diagrams in
Fig.~\ref{annihilation}) contributions, respectively.}
\begin{center}
\doublerulesep 0.8pt \tabcolsep 0.15in
\begin{tabular}{lcccc}\hline \hline
Decay modes & $A_{fac}$&$A_{nonfac}$&$A_{anni}$\\
 \hline\hline
 $\overline B^{0} \to D^{+}\pi^-$ &$- 2.2655-0.0060i$
 & $0.2613- 0.1862i$&$-0.0047+ 0.0053i$
 \\$\overline B^{0}\to D^{-}\pi^+$&$1.7865-0.0556i$&$-0.0447+0.0970i$
 &$0.0008-0.0017i$\\
 $\overline B^{0} \to D^0\pi^{0}$&$-0.0603+0.0011i$&
 $0.6991 -0.0022i$&$0.0010+ 0.0009i$\\
 $ B^{-}\to D^{0} \pi^{-}$ &$-2.3375-0.2310i$&
 $-0.4884-0.1484i$&$0$\\
  \hline\hline
\end{tabular}
\end{center}
\end{table}

As a consequence, we are led to the predictions for the quantities
$z$ and $\delta$, as well as the branching ratios of all the $B\to
D\pi$ decay modes.  We present our ``default results" of branching
ratios and detailed error estimates corresponding to the different
theoretical uncertainties caused from the above input parameters in
Tables~\ref{br1} and ~\ref{br2}, respectively. The errors consist of
three parts: the first one refers to the variation of the dynamical
gluon mass scale; the second one arises from the uncertainty due to
the CKM parameters $A, \lambda, \bar{\rho}$, and $\bar{\eta}$; the
third one originates from the uncertainties due to the meson decay
constants and the parameter $\mu_\pi$.

\begin{table}[t]
\caption{\label{br1} The branching ratios of $B \rightarrow D\pi$
decays with the default input parameters. The theoretical results in
the second and the third lines correspond to the predictions at the
$m_b/2$ and $m_b$ scales, respectively. The results correspond to
$\mu_g=440~{\rm MeV}$.}
\begin{center}
\doublerulesep 0.8pt \tabcolsep 0.1in
\begin{tabular}{lcccccc}\hline \hline
Decay modes &$Br(m_b/2)$ &$Br(m_b)$ &Experiment & Ref.~\cite{Keum:2003js} \\
\hline\hline
 $\overline B^{0} \to D^{+}\pi^-(10^{-3})$ & $2.67$&$2.20$
 & $2.68\pm0.12\pm0.24\pm0.12$~\cite{vonToerne:2003gc}& $2.74^{+0.39}_{-0.37}$
 \\$$&$$ &$$&$2.63\pm0.05\pm0.22$~\cite{Aubert:2006cd}& $$
 \\\hline $ B^{-}\to D^{0} \pi^{-}(10^{-3})$ &$4.87$&$4.73$
 &$4.97\pm0.12\pm0.29\pm0.22$~\cite{vonToerne:2003gc}
 & $5.43^{+0.48}_{-0.47}$\\
 $$ &$$&$$ &$4.90\pm0.07\pm0.23$~~\cite{Aubert:2006cd}
 & $$\\\hline
 $\overline B^{0} \to D^0\pi^{0}(10^{-4})$ &$2.17$ &$2.04$&$2.89\pm0.29\pm0.38$
 ~\cite{Aubert:2002yf} & $2.5\pm0.1$\\
 $$&$$ &$$&
 $2.25\pm0.14\pm0.35$\cite{Blyth:2006at} & $$\\\hline
 $\overline B^{0}\to D^{-}\pi^+(10^{-7})$  &$5.54$& $4.78$&$-$&$-$\\
  \hline\hline
\end{tabular}
\end{center}
\end{table}
\begin{table}[t]
\caption{\label{br2} The branching ratios of $B \rightarrow D\pi$
decays. The theoretical errors shown from left to right correspond
to the uncertainties referred to as ``dynamical gluon mass
scale(upper one corresponding to $\mu_g=420~{\rm MeV}$ and the below
one $\mu_g=460~{\rm MeV}$)", ``CKM parameters", and ``decay constants
and the parameter $\mu_{\pi}$" as specified in the text. }
\begin{center}
\doublerulesep 0.8pt \tabcolsep 0.3in
\begin{tabular}{lcccc}\hline \hline
Decay modes &$Br(m_b/2)$ &$Br(m_b)$ \\
\hline\hline
 $\overline B^{0} \to D^{+}\pi^-(10^{-3})$ & $2.67^{\,+0.47\,+0.62\,+0.91}
  _{\,-0.36\,-0.57\,-0.85}$&$2.20^{\,+0.52\,+0.16
 \,+0.80} _{\,-0.51\,-0.42\,-0.62}$
  \\ $ B^{-}\to D^{0} \pi^{-}(10^{-3})$ &$4.87^{\,+0.52\,+0.84\,+1.74}
 _{\,-0.48\,-0.50\,-0.31}$&$4.73^{\,+0.63\,+0.34\,+1.29}
 _{\,-0.44\,-0.89\,-1.18}$ \\
 $\overline B^{0} \to D^0\pi^{0}(10^{-4})$ &$2.17^{\,+0.53\,+0.46\,+1.14}
 _{\,-0.45\,-0.46\,-0.89}$ &$2.04^{\,+0.56\,+0.19\,+0.95}
  _{\,-0.33\,-0.50\,-0.75}$\\
 $\overline B^{0}\to D^{-}\pi^+(10^{-7})$  &$5.54^{\,+0.68\,+1.02\,+1.52}
 _{\,-0.44\,-0.97\,-1.38}$& $4.78^{\,+0.62\,+0.81\,+1.76}
  _{\,-0.48\,-1.10\,-1.38}$ \\\hline\hline
\end{tabular}
\end{center}
\end{table}

From the numerical results given in Tables~\ref{amplitude}-
\ref{br2}, we arrive at the following observations:

(i) For the color-allowed (also Cabibbo-favored) decay mode
$\overline B^0 \rightarrow D^+\pi^-$, the factorizable contribution
$A_{fac}$ dominates the total decay amplitude, while the
contributions from $A_{nonfac}$ and $A_{anni}$ are small. In
particular, the contribution of $A_{anni}$ is so small that we can
safely neglect it in this decay mode. With the considered
uncertainties, it is seen that our result is in agreement with the
experimental data~\cite{vonToerne:2003gc,Aubert:2006cd}, and also
consistent with the one given in~\cite{Keum:2003js}: ${\cal
B}(\overline B^0\rightarrow D^+\pi^-)=(2.74^{+0.39}_{-0.37})\times
10^{-3}$ within the allowed theoretical uncertainties. The decay
amplitude of its $CP$-conjugate decay mode $B^0\rightarrow D^-\pi^+$
can be obtained from that of $\overline B^0 \rightarrow D^+\pi^-$ by
changing the CKM element $V_{cb}V^*_{ud}$ to $V^*_{cb}V_{ud}$. Since
these CKM elements are purely real, the branching ratio of
$B^0\rightarrow D^-\pi^+$ decay is the same as that of $\overline
B^0 \rightarrow D^+\pi^-$ decay.

(ii) For the doubly Cabibbo-suppressed decay mode $\overline B^0
\rightarrow D^-\pi^+$, the contributions from $A_{nonfac}$ and
$A_{anni}$ are also much smaller than the one from $A_{fac}$. As the
contributions are all proportional to the small CKM elements
$|V_{ub}V^*_{cd}|$ $\sim \lambda^4$, the branching ratio of this
decay mode is found to be at the order of $10^{-7}$, and much
smaller than that of the Cabibbo-favored decay mode. Since the
imaginary part of the dominated amplitude $A_{fac}$ is much smaller
than the real part, the branching ratio of the $CP$-conjugate decay
mode $B^0 \rightarrow D^+\pi^-$ is approximately equal to that of
the $\overline B^0 \rightarrow D^-\pi^+$ decay.

(iii) For the color-suppressed decay mode $\overline B^0\to
D^0\pi^0$, the contribution from $A_{nonfac}$ dominates the total
decay amplitude. The result at the $m_b$ scale is smaller than that
at the $m_b/2$ scale, but both are in agreement with the prediction
given in~\cite{Keum:2003js}: ${\cal B}(\overline B^0\rightarrow
D^0\pi^0)=(2.5\pm 0.1)\times 10^{-4}$. The present central value at
the $m_b/2$ scale agree well with the experimental data reported
in~\cite{ Blyth:2006at}, but slightly smaller than the recent
experimental data given in~\cite{Aubert:2002yf}. While when
considering their respective uncertainties, our prediction is still
consistent with the experimental data.

(iv) For the $B^-\to D^0\pi^-$ decay, the main contribution
originates from the factorizable one $A_{fac}$. Although its decay
amplitude can be written as the color-favored $\overline
B^0\rightarrow D^+\pi^-$ minus the color suppressed $\overline
B^0\rightarrow D^0\pi^0$ decays, the branching ratio of this decay
mode is enhanced compared to that of the latter two. The central
values of our prediction are well consistent with the experimental
data given in~\cite{Aubert:2006cd,vonToerne:2003gc}. On the other
hand, when taking into account of the theoretical uncertainties, our
prediction is also consistent with the one given by the pQCD
method~\cite{Keum:2003js}: ${\cal B}(B^-\to
D^0\pi^-)=(5.43^{+0.48}_{-0.47})\times 10^{-3}$.

(v) Although the branching ratios at the $m_b$ scale are smaller
than those at the $m_b/2$ scale in all these decay modes, we can see
that the final results have only a marginal dependence on the
renormalization scale. As for the theoretical uncertainties in these
decay modes, the errors originating from the dynamical gluon mass
scale $\mu_g$, the CKM matrix elements are comparable with each
other when $\mu_g\in(420, 460)~{\rm MeV}$. However, the uncertainty
originating from the decay constants and $\mu_\pi$ are dominate in
these decays, especially in $B^-\to D^0\pi^-$ and $\overline B^0\to
D^-\pi^+$ modes.

It is also interesting to note that the $B\to D$ transition form
factor, $F^{B\to D}=0.634$, extracted from (a) and (b) diagrams in
Fig.~\ref{tree} is in good agreement with the ones obtained from the
other methods, such as: $F^{B\to D}(0)=0.648$ (pQCD
method~\cite{Kurimoto:2002sb}); $F^{B\to D}(0)=0.690$
(Bauer-Stech-Wirbel(BSW) model~\cite{Wirbel}); $F^{B\to D}(0)=0.636$
(Neubert-Stech(NS) model~\cite{Neubert:1997uc}).

We now turn to discuss the $CP$ asymmetries in $B\to D\pi$ decays.
As has already been discussed above, there are no direct $CP$
violations in all these decay modes. In the following discussions,
we focus mainly on the time-dependent $CP$ asymmetries of $B\to
D^\pm\pi^\mp$ decays.

\begin{figure}[hbpt]
\begin{center}
\begin{tabular}{cc}
\scalebox{0.9}{\epsfig{file=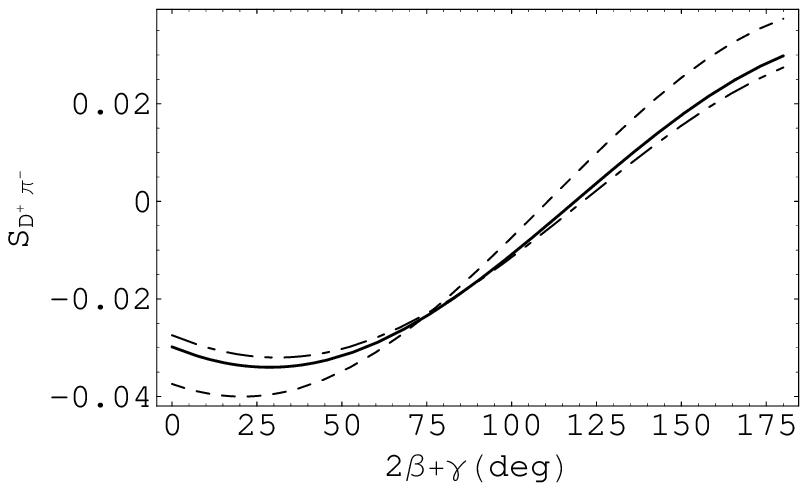}}& \scalebox{0.9}
{\epsfig{file=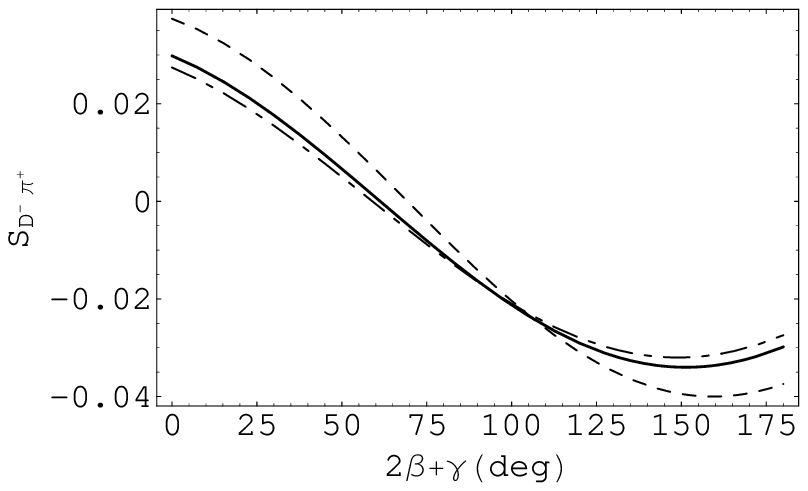}}
\end{tabular}
\caption{\label{cpv1} \small The $CP$-violating parameters
$S_{D^+\pi^-}$ and $S_{D^-\pi^+}$ for $B \to D^\pm \pi^\mp$ decays
as functions of the weak phase $2\beta+\gamma$~(in degree). The
dashed, solid, and dash-dotted lines correspond to $\mu_g=1.3, 1.5$
and $1.7\Lambda_{\rm QCD}$, respectively.}
\end{center}
\end{figure}

\begin{figure}[hbpt]
\begin{center}
\begin{tabular}{cc}
\scalebox{0.9}{\epsfig{file=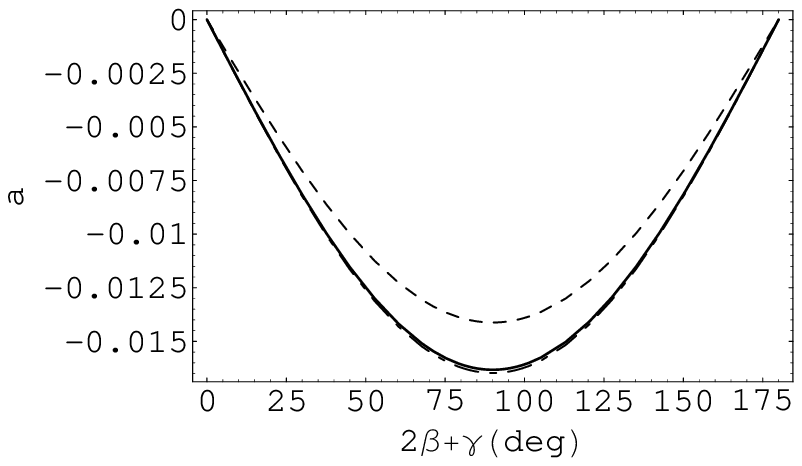}}& \scalebox{0.9}
{\epsfig{file=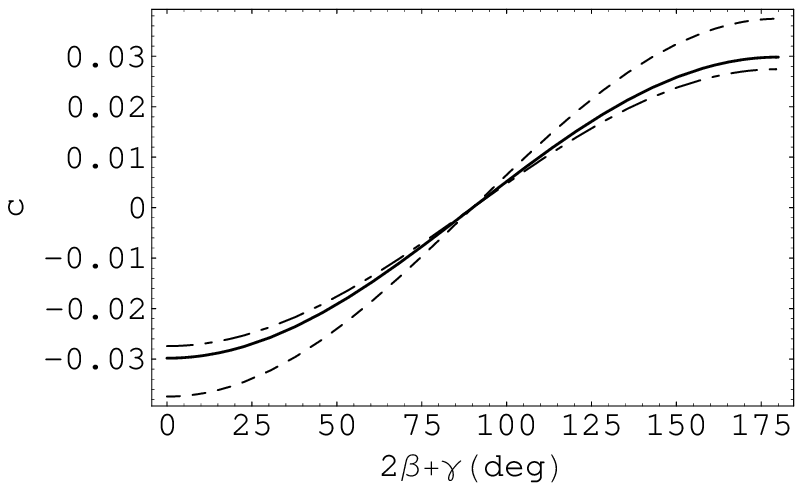}}
\end{tabular}
\caption{\label{cpv2} \small The same as Fig.~\ref{cpv1}, but for
$CP$ observables $a$ and $c$.}
\end{center}
\end{figure}

Using the relevant formulas presented in the previous sections, we
can predict the $CP$ asymmetries in $B\to D^\pm\pi^\mp$ decays and
constrain the CKM angle $2\beta+\gamma$ through the two observables
$a$ and $c$. Firstly, we present the results of the quantities $z$
and $\delta$ in Table~\ref{cp}. Taking the current constraints for
the weak angles $\beta$ and $\gamma$ in the SM, we present our
predictions for the $CP$ asymmetries $S_{D^+\pi^-}$ and
$S_{D^-\pi^+}$, as well as the two observables $a$ and $c$.
Secondly, taking the weak angles $\beta$ and $\gamma$ as free
parameters, we show the dependence of the parameters $S_{D^+\pi^-},
S_{D^-\pi^+}$ and the observables $a$, $c$ on the angle
$2\beta+\gamma$ in Figs.~\ref{cpv1} and \ref{cpv2}, respectively.

\begin{table}[hbpt]
\caption{\label{cp} \small The $CP$ asymmetries for $B \to D^\pm
\pi^\mp$ decays. The results in the middle row denote our
theoretical predictions for each quantity. The center values
correspond to $\mu_g=1.5\Lambda_{\rm QCD}$, and the error bars
originate from the dynamical gluon mass scale with the upper one
corresponding to $\mu_g=1.3\Lambda_{\rm QCD}$ and the below one
$\mu_g=1.7\Lambda_{\rm QCD}$. The results in last row are the
experimental data.}
\begin{center}
\doublerulesep 0.9pt \tabcolsep 0.05in
\begin{tabular}{lccccccc}\hline \hline
  results & $z$&$\delta$&$S_{D^+\pi^-}$&S$_{D^-\pi^+}$&$a$&$c$
\\\hline\hline
 Theor&$0.017^{+0.03}_{-0.01}$ &$1.07^{+0.14}_{-0.04}$&
  $-0.010^{+0.004}_{-0.001}$&$-0.023^{+0.001}_{-0.001}$&$-0.017^{+0.003}
  _{-0.001}$&$0.007^{+0.002}_{-0.001}$\\
Exp~\cite{Barberio:2006bi}
&$-$&$-$&$-$&$-$&$-0.030\pm0.017$&$-0.022\pm0.021$
 \\\hline\hline
\end{tabular}
\end{center}
\end{table}

From Table~\ref{cp} and Figs.~\ref{cpv1} and \ref{cpv2}, we come to
the following observations.
\begin{itemize}
\item[]{(i)} For $B\to D^\pm\pi^\mp$ decay modes, although there are
large strong phase difference between the Cabibbo-suppressed and the
Cabibbo-favored decay amplitudes, the $CP$-violating parameters
$S_{D^+\pi^-}$ and $S_{D^-\pi^+}$ are found to be
small~($-0.01\sim-0.02$) due to the smallness of the ratio $z$~(
$\simeq0.02$). In addition, due to our predictions for the two
parameters $S_{D^+\pi^-}$ and $S_{D^-\pi^+}$ are comparable to each
other for a given dynamical gluon mass scale, the value for the
parameter $c$ is nearly zero.

\item[]{(ii)} The $CP$-violating parameters $S_{D^+\pi^-}$ and $S_{D^-\pi^+}$
are not sensitive to the choice of the dynamical gluon mass scale,
especially when the dynamical gluon mass scale is chosen above the
central value $1.5\Lambda_{\rm QCD}$. However, both of them have a
strong dependence on the weak angle $2\beta+\gamma$. The same
conclusion is also applied to the two observables $a$ and $c$. Our
predictions for the two observables are consistent with experimental
data when considering the corresponding uncertainties.
Unfortunately, it is found that with the angle $2\beta+\gamma$
varying within the range $(0,180^{\circ})$, almost all of the values
for $a$ and $c$ are in the range of the experimental data, which
indicates that although direct constraints on the angle
$2\beta+\gamma$ could be obtained through these parameters, the
present experimental accuracy is insufficient to improve the
knowledge of the apex in the unitarity plane. It is expected that
more precise measurements in future experiments allow us to extract
the angle $2\beta+\gamma$.
\end{itemize}


\section{Conclusions}

In summary, we have calculated the decay amplitudes, strong phases,
branching ratios, and $CP$ asymmetries for the $B\to D\pi$ decays,
including both the color-allowed and the color-suppressed modes. It
has been shown that these decay modes are theoretically clean as
there are no penguin contributions. As a consequence, direct $CP$
violations are absent. The contributions from the factorizable
diagrams dominate all the decay amplitudes except for the $\overline
B^0\to D^0\pi^0$ process. All our predictions for branching ratios
are consistent with the existing measurements. For the $\overline
B^0\to D^-\pi^+$ mode, our predictions will be faced with the future
experiments as no data are available at present. Due to small
interference effects between the Cabibbo-suppressed and the
Cabibbo-favored amplitudes, the non-zero $CP$-violating parameters
$S_{D^+\pi^-}$ and $S_{D^-\pi^+}$ have been predicted in the $B \to
D^\pm\pi^\mp$ decay modes. It has been shown that the $CP$-violating
parameters have a strong dependence on the weak phase
$2\beta+\gamma$, but they are not sensitive to the dynamical gluon
mass scale. With the angle $2\beta+\gamma$ varying within the range
$(0,180^{\circ})$, almost all of the values for the CP-violating
parameters $a$ and $c$ are within the range of the current
experimental data. Thus no constraints on the weak phase
$2\beta+\gamma$ could be obtained through those parameters based on
the current experiment data, and more precise measurements are
needed in future experiments.

In this paper, we have further shown that the divergence treatments
used in our previous work~\cite{kk} are reliable. Namely, the
endpoint divergence caused by the soft collinear approximation in
gluon propagator could be simply avoided by adopting the Cornwall
prescription for the gluon propagator with a dynamical mass scale.
Note that when the intrinsic mass is appropriately introduced, it
may not spoil the gauge symmetry as shown recently in the
symmetry-preserving loop regularization~\cite{LR}. Meanwhile, for
the physical-region singularity of the on-mass-shell quark
propagators, it can well be treated by applying for the Cutkosky
rules. The combination of these two treatments for the endpoint
divergences of gluon propagator and the physical-region singularity
of the quark propagators enables us to obtain reasonable results,
which are consistent with the existing experimental data and also in
agreement with the ones~\cite{Keum:2003js} obtained by using the
pQCD approach. However, this is different from the treatment of the
latter, where $k_{T}^{2}$ and  Sudakov factors have been used to avoid the
endpoint divergence.

It is noted that the resulting predictions for the branching ratios
are in general scale dependence on the dynamical gluon mass which
plays the role of the IR cut-off. This dependence should in
principle be compensated from the possible scale in the wave
functions which characterizes the nonperturbative effects. In our
approach, the dynamical gluon mass may be regarded as a universal
scale to be fixed from one of the decay modes. For instance, in our
present considerations, if the decay mode $\bar{B}^0 \to D^+\pi^-$
is taken to extract the dynamical gluon mass scale, we have $\mu_g
\simeq 440$ MeV, and the resulting predictions for other decay modes
can serve as a consistent check. Within the current experimental
errors and theoretical uncertainties for some relevant parameters,
it is seen that our treatment is reliable. In order to further check
the validity of the gluon-mass regulator method adopted to deal with
the endpoint divergence, it is useful to extend this method to more
decay modes. Anyway, the treatments presented in this paper may
enhance its predictive power for analyzing non-leptonic $B$-meson
decays.

\acknowledgments This work was supported in part by the National
Science Foundation of China (NSFC) under the grant 10475105,
10491306, 10675039 and the Project of Knowledge Innovation Program
(PKIP) of Chinese Academy of Sciences.

\section*{Appendix: Detail calculations of the $B\to D\pi$ decays amplitudes}

To evaluate the hadronic matrix elements of $B\to D\pi$ decays, the
meson light-cone distribution amplitudes play an important role. In
the heavy quark limit, the light-cone projectors for $B$, $D$ and
$\pi$ mesons in momentum space can be expressed, respectively,
as~\cite{M.T}
\begin{eqnarray}
 {\cal M}_{\alpha\beta}^B &=& -\frac{i f_B }{4}\,
 [(m_B+\spur{P_1} )\,\gamma_5 \,\phi_B(\rho)]_{\alpha\beta}, \nonumber\\
 {\cal M}_{\alpha\beta}^D &=& \frac{i f_D\,}{4}\,
 [(\spur{P_2}+ m_D )\,\gamma_5 \,\phi_D(y)]_{\alpha\beta}, \nonumber\\
{\cal M}_{\delta\alpha}^\pi &=& \frac{i
f_P}{4}\,\biggl\{\spur{P_3}\gamma_5 \,\phi(u)-\mu_P\gamma_5
\biggl(\phi_p(u)-i\sigma_{\mu \nu}n_-^\mu v^\nu \frac{\phi^{\prime}_
\sigma (u)}{6} + i \sigma _{\mu \nu} P_3^{\mu}
\frac{\phi_\sigma(u)}{6} \frac{\partial}{\partial
k_{\bot\nu}}\biggl)\biggl\}_{\delta\alpha},\label{projector}
 \end{eqnarray}
From the Feynman diagrams shown in
Figs.~\ref{tree}-\ref{annihilation}, we can get the amplitudes for
each decay mode using the relevant Feynman rules and the light-cone
projectors listed in Eqs.~(\ref{projector}).

For the tree diagrams of $\overline B^0\to D^+\pi^-$ mode shown in
Fig.~\ref{tree}, the amplitudes of each diagrams can be written as
\begin{eqnarray}
A^{\ref{tree}a}&=&if_\pi P_3^\mu \mathrm{Tr}\big[{\cal
M}^B(-ig_s\gamma^\alpha T^a_{ij}){\cal M}^D\gamma_\mu
(1-\gamma_5)\frac{i}{\spur{k_b}-m_b}(-ig_s\gamma^\beta
T^b_{kl})\big]\frac{-ig_{\alpha\beta}\delta_{ab}}{k^2},\nonumber\\
&=&-i f_\pi g^2_s\frac{C_F}{N_C}\frac{1}{D_b
k^2}\mathrm{Tr}\big[{\cal M}^B\gamma^\alpha{\cal M}^D\spur{P_3}
(1-\gamma_5)(\spur{k_b}+m_b)\gamma_\alpha\big]\nonumber\\
A^{\ref{tree}b}&=&if_\pi P_3^\mu\mathrm{Tr}\big[{\cal
M}^B(-ig_s\gamma^\alpha T^a_{ij}){\cal M}^D(-ig_s\gamma^\beta
T^b_{kl})\frac{i}{\spur{k_c}-m_c}\gamma_\mu
(1-\gamma_5)\big]\frac{-ig_{\alpha\beta}\delta_{ab}}{k^2},\nonumber\\
&=&-if_\pi g^2_s\frac{C_F}{N_C}\frac{1}{D_c
k^2}\mathrm{Tr}\big[{\cal M}^B\gamma^\alpha{\cal
M}^D\gamma_\alpha(\spur{k_c}+m_c)\spur{P_3}
(1-\gamma_5)\big]\nonumber\\
A^{\ref{tree}c}&=&\mathrm{Tr}\big[{\cal M}^\pi(-ig_s\gamma^\alpha
T^a_{ij})\frac{i}{\spur{k_d}}\gamma^\mu
(1-\gamma_5)\big]\mathrm{Tr}\big[{\cal M}^B(-ig_s\gamma^\beta
T^b_{kl}){\cal
M}^D\gamma_\mu (1-\gamma_5)\big]\frac{-ig_{\alpha\beta}\delta_{ab}}{k^2},\nonumber\\
&=&-g^2_s\frac{C_F}{N_C}\frac{1}{D_d k^2}\mathrm{Tr}\big[{\cal
M}^\pi\gamma^\alpha\spur{k_d}\gamma^\mu
(1-\gamma_5)\big]\mathrm{Tr}\big[{\cal M}^B\gamma_\alpha{\cal
M}^D\gamma_\mu (1-\gamma_5)\big]\nonumber\\
A^{\ref{tree}d}&=&\mathrm{Tr}\big[{\cal M}^\pi\gamma^\mu
(1-\gamma_5)\frac{i}{\spur{k_u}}(-ig_s\gamma^\alpha
T^a_{ij})\big]\mathrm{Tr}\big[{\cal M}^B(-ig_s\gamma^\beta
T^b_{kl}){\cal M}^D\gamma_\mu
(1-\gamma_5)\big]\frac{-ig_{\alpha\beta}\delta_{ab}}{k^2}\nonumber\\
&=&-g^2_s\frac{C_F}{N_C}\frac{1}{D_u k^2}\mathrm{Tr}\big[{\cal
M}^\pi\gamma^\mu (1-\gamma_5)\spur{k_u}\gamma^\alpha
\big]\mathrm{Tr}\big[{\cal M}^B\gamma_\alpha{\cal M}^D\gamma_\mu
(1-\gamma_5)\big],\label{amplitude1}
\end{eqnarray}
where $A^{\ref{tree}j}$ stands for the $j$th($j=a,b,c,d$) diagrams
in Fig.\ref{tree}, $k_m$ and $k$ the momentum of $m$ quark
propagator and gluon propagator, respectively. Furthermore, $D_m$
and $k^2$ represent for the $m$ quark propagator and gluon
propagator, respectively.

In Fig.~\ref{tree}(a), the $\pi$ meson can be written as a decay
constant since it originates from the vacuum. Inversing the fermi
lines and writing down the $B$ meson projector ${\cal M}^B$, gluon
vertex $-ig_s\gamma^\alpha T^a_{ij}$, $D$ meson projector ${\cal
M}^D$, the four quark vertex $\gamma_\mu(1-\gamma_5)$, b quark
propagator $\frac{i}{\spur{k}_b-m_b}$ and another gluon vertex
$-ig_s\gamma^\beta T^b_{kl}$ in a trace one by one, and finally the
gluon propagator $\frac{-ig_{\alpha\beta}\delta_{ab}}{k^2}$, we can
get the amplitude $A^{\ref{tree}a}$. $A^{\ref{tree}b}$ can be
calculated in a similar way. In Fig.~\ref{tree}(c), the $\pi$ meson
can no longer be written as a decay constant any more since it
exchanges a gluon with the spectator quark. Writing down the $\pi$
meson projector ${\cal M}^\pi$, gluon vertex $-ig_s\gamma^\alpha
T^b_{ij}$, $d$ quark propagator $\frac{i}{\spur{k}_d}$ and the four
quark vertex $\gamma^\mu(1-\gamma_5)$ in turn in one trace, and
writing down the $B$ meson projector ${\cal M}^B$, gluon vertex
$-ig_s\gamma^\beta T^b_{kl}$, $D$ meson projector ${\cal M}^D$ and
the four quark vertex $\gamma_\mu(1-\gamma_5)$ in the other trace
one by one, and finally the gluon propagator
$\frac{-ig_{\alpha\beta}\delta_{ab}}{k^2}$, we can get the amplitude
$A^{\ref{tree}c}$. Similarly, we can get the amplitude
$A^{\ref{tree}d}$. Summing up the former and the latter two
quantities in Eq.~(\ref{amplitude1}), we can get the factorizable
part $A_{fac}$~(Eq.~(\ref{favor2})) and the nonfactorizable
$A_{nonfac}$~(Eq.~(\ref{favor3})), respectively.

 As for the annihilation diagrams for $\overline B^0 \to D^+\pi^-$ in
Fig.~\ref{annihilation}, the amplitudes can be written as
\begin{eqnarray}
A^{\ref{annihilation}a}&=&if_B P_1^\mu \mathrm{Tr}\big[{\cal
M}^D(-ig_s\gamma^\alpha
T^a_{ij})\frac{i}{\spur{k_{ca}}-m_c}\gamma_\mu (1-\gamma_5){\cal
M}^\pi(-ig_s\gamma^\beta
T^b_{kl})\big]\frac{-ig_{\alpha\beta}\delta_{ab}}
{k^2_a},\nonumber\\
&=&-i f_B g^2_s\frac{C_F}{N_C}\frac{1}{D_{ca}
k^2_a}\mathrm{Tr}\big[{\cal M}^D\gamma^\alpha
(\spur{k_{ca}}+m_c)\spur{P_1}(1-\gamma_5){\cal
M}^\pi\gamma_\alpha\big]\nonumber\\
A^{\ref{annihilation}b}&=&if_B P_1^\mu \mathrm{Tr}\big[{\cal
M}^D\gamma_\mu
(1-\gamma_5)\frac{i}{\spur{k_{ua}}}(-ig_s\gamma^\alpha
T^a_{ij}){\cal M}^\pi(-ig_s\gamma^\beta
T^b_{kl})\big]\frac{-ig_{\alpha\beta}\delta_{ab}}{k^2_a},\nonumber\\
&=&-i f_B g^2_s\frac{C_F}{N_C}\frac{1}{D_{ua}
k^2_a}\mathrm{Tr}\big[{\cal M}^D\spur{P_1}
(1-\gamma_5)\spur{k_{ua}}\gamma^\alpha{\cal M}^\pi\gamma_\alpha\big],\nonumber\\
A^{\ref{annihilation}c}&=&\mathrm{Tr}\big[{\cal
M}^B(-ig_s\gamma^\alpha T^a_{ij})\frac{i}{\spur{k_{da}}}\gamma^\mu
(1-\gamma_5)\big]\mathrm{Tr}\big[{\cal M}^D\gamma_\mu
(1-\gamma_5){\cal M}^\pi(-ig_s\gamma^\beta
T^b_{kl})\big]\frac{-ig_{\alpha\beta}\delta_{ab}}{k^2_a}\nonumber\\
&=&- g^2_s\frac{C_F}{N_C}\frac{1}{D_{da} k^2_a}\mathrm{Tr}\big[{\cal
M}^B\gamma^\alpha \spur{k_{da}}\gamma^\mu
(1-\gamma_5)\big]\mathrm{Tr}\big[{\cal M}^D\gamma_\mu
(1-\gamma_5){\cal M}^\pi\gamma_\alpha \big],\nonumber\\
A^{\ref{annihilation}d}&=&\mathrm{Tr}\big[{\cal M}^B\gamma^\mu
(1-\gamma_5)\frac{i}{\spur{k_{ba}-m_b}}(-ig_s\gamma^\alpha
T^a_{ij})\big]\mathrm{Tr}\big[{\cal M}^D\gamma_\mu (1-\gamma_5){\cal
M}^\pi(-ig_s\gamma^\beta
T^b_{kl})\big]\frac{-ig_{\alpha\beta}\delta_{ab}}
{k^2_a},\nonumber\\
&=&- g^2_s\frac{C_F}{N_C}\frac{1}{D_{ba} k^2_a}\mathrm{Tr}\big[{\cal
M}^B\gamma^\mu (1-\gamma_5)(\spur{k_{ba}}+m_b)\gamma^\alpha
\big]\mathrm{Tr}\big[{\cal M}^D\gamma_\mu (1-\gamma_5){\cal
M}^\pi\gamma_\alpha \big],\label{amplitude1a}
\end{eqnarray}
where $k_{ma}$ and $k_a$ stand for the momentum of $m$ quark
propagator and gluon propagator, and $D_{ma}$ and $k^2$ represent
for the $m$ quark propagator and gluon propagator in these
annihilation diagrams, respectively. Summing up the four quantities
in Eq.~(\ref{amplitude1a}), we can get the annihilation contribution
$A_{anni}$~(Eq.~(\ref{anni1})) of this decay mode.

 Similarly, as for the tree diagrams of $\overline B^0 \to D^0\pi^0$ decay mode
 in Fig~\ref{tree2}, its amplitudes can be written as follows
\begin{eqnarray}
A^{\ref{tree2}a}&=&if_D P_2^\mu \mathrm{Tr}\big[{\cal
M}^B(-ig_s\gamma^\alpha T^a_{ij}){\cal M}^\pi\gamma_\mu
(1-\gamma_5)\frac{i}{\spur{k_b}-m_b}(-ig_s\gamma^\beta
T^b_{kl})\big]\frac{-ig_{\alpha\beta}\delta_{ab}}{k^2},\nonumber\\
&=& -if_D g^2_s\frac{C_F}{N_C}\frac{1}{D_{b}
k^2}\mathrm{Tr}\big[{\cal M}^B\gamma^\alpha{\cal M}^\pi\spur{P_2}
(1-\gamma_5)(\spur{k_b}+m_b)\gamma_\alpha\big],\nonumber\\
A^{\ref{tree2}b}&=&if_D P_2^\mu \mathrm{Tr}\big[{\cal
M}^B(-ig_s\gamma^\alpha T^a_{ij}){\cal M}^\pi(-ig_s\gamma^\beta
T^b_{kl})\frac{i}{\spur{k_d}}\gamma_\mu
(1-\gamma_5)\big]\frac{-ig_{\alpha\beta}\delta_{ab}}{k^2},\nonumber\\
&=& -if_D g^2_s\frac{C_F}{N_C}\frac{1}{D_{d}
k^2}\mathrm{Tr}\big[{\cal M}^B\gamma^\alpha{\cal M}^\pi\gamma_\alpha
\spur{k_d}\spur{P_2}(1-\gamma_5)\big],\nonumber\\
A^{\ref{tree2}c}&=&\mathrm{Tr}\big[{\cal M}^D(-ig_s\gamma^\alpha
T^a_{ij})\frac{i}{(\spur{k_c}-m_c)}\gamma^\mu
(1-\gamma_5)\big]\mathrm{Tr}\big[{\cal M}^B(-ig_s\gamma^\beta
T^b_{kl}){\cal M}^\pi\gamma_\mu (1-\gamma_5)\big]
\frac{-ig_{\alpha\beta}\delta_{ab}}{k^2},\nonumber\\
&=& -ig^2_s\frac{C_F}{N_C}\frac{1}{D_{c} k^2}\mathrm{Tr}\big[{\cal
M}^D\gamma^\alpha(\spur{k_c}+m_c)\gamma^\mu
(1-\gamma_5)\big]\mathrm{Tr}\big[{\cal M}^B\gamma_\alpha{\cal
M}^\pi\gamma_\mu (1-\gamma_5)\big],\nonumber\\
A^{\ref{tree2}d}&=&\mathrm{Tr}\big[{\cal M}^D\gamma^\mu
(1-\gamma_5)\frac{i}{\spur{k_u}}(-ig_s\gamma^\alpha
T^a_{ij})\big]\mathrm{Tr}\big[{\cal M}^B(-ig_s\gamma^\beta
T^b_{kl}){\cal M}^\pi\gamma_\mu
(1-\gamma_5)\big]\frac{-ig_{\alpha\beta}\delta_{ab}}{k^2}\nonumber\\
&=& -ig^2_s\frac{C_F}{N_C}\frac{1}{D_{u} k^2}\mathrm{Tr}\big[{\cal
M}^D\gamma^\mu (1-\gamma_5)\spur{k_u}\gamma^\alpha
\big]\mathrm{Tr}\big[{\cal M}^B\gamma_\alpha{\cal M}^\pi\gamma_\mu
(1-\gamma_5)\big].\label{amplitude2}
\end{eqnarray}
We can get the factorizable contribution $A_{fac}$
(Eq.~(\ref{ckms1})) and the nonfactorizable part
$A_{nonfac}$~(Eq.~(\ref{ckms2})) by summing up the formerand the
latter two quantities in Eq.~(\ref{amplitude2}).

As for the annihilation diagrams for $\overline B^0\to D^0\pi^0$
decay, its amplitude is the same as the one in
Eq.~(\ref{amplitude1a}) since the two modes $D^0\pi^0$ and
$D^+\pi^-$ have the same annihilation topological diagrams, which
are shown in Fig.~\ref{annihilation}.

For the doubly Cabibbo-suppressed decay mode $\overline B^0\to
D^-\pi^+$, its decay amplitude can be similarly expressed as the
ones in Eq.~(\ref{amplitude2}) due to the same topological structure
in these two decay modes.

Finally, for the $B^-\to D^0\pi^-$ decay, since its Feynman diagrams
are the sums of the color-allowed and the color-suppressed one, we
can easily get its amplitudes using the above results.


\end{document}